# Density Functional Theory Study of Controllable Optical Absorptions and Magneto-Optical Properties of Magnetic CrI$_3$ Nanoribbons: Implications for Compact 2D Magnetic Devices

Hong Tang[1], Santosh Neupane, Qimin Yan, and Adrienn Ruzsinszky

Department of Physics, Temple University, Philadelphia, PA 19122

**ABSTRACT**   A chromium triiodide (CrI$_3$) monolayer has an interesting ferromagnetic ground-state. In this work, we calculate band structures and magnetic moments of tensile-strained and bent zigzag CrI$_3$ nanoribbons with density functional theory. The edge iodine atoms form flat low-lying conduction bands and couple with chromium atoms ferromagnetically, while the non-edge iodine atoms weakly couple antiferromagnetically. Narrow CrI$_3$ nanoribbons have two locally stable magnetic moment orientations, namely out-of-plane and in-plane (along the nanoribbon periodic direction) configurations. This enables four magnetization states in CrI$_3$ nanoribbons, including two out-of-plane ones (up and down) and two in-plane ones (forward and backward along the nanoribbon periodical direction), increasing the operating controllability. Based on the one-dimensional Ising spin chain model, the spin correlation length of the narrow CrI$_3$ nanoribbon is estimated as about 10 Å at its estimated Curie temperature of 27 Kelvin, which is lower than the measured 45 Kelvin of the monolayer CrI$_3$. The optical absorption and magneto-optical properties of CrI$_3$ nanoribbons are investigated with many-body perturbation GW-BSE (Bethe-Salpeter equation), including magnetic dichroism, Faraday and magneto-optical Kerr effects. The low-energy dark excitons are mainly from transitions between electrons and holes with unlike spins and are non-Frenkel-like, while the bright excitons have mixed spin configurations. The intrinsic lifetime of excitons can be over one nanosecond, suitable for quantum information processes. Tensile strains and bending manifestly modulate the absorption spectra and magneto-optical properties of CrI$_3$ nanoribbons within a technologically important photon energy-range of ~1.0-2.0 eV. The CrI$_3$ nanoribbons can be used in 1D or 2D magnetic storage nanodevices, tunable magnetic optoelectronics, and spin-based quantum information controls.

## INTRODUCTION

The discovery of intrinsic ferromagnetism (FM) in a monolayer of the two-dimensional (2D) layered material chromium triiodide (CrI$_3$) is interesting,[1] since the Mermin-Wagner theorem[2] theoretically predicts that no magnetic order can exist in the 2D isotropic Heisenberg model at finite temperatures. However, one can realize the long-range FM order in 2D systems by breaking the isotropic spin-rotational invariance.[3] From the measurement of the zero wavevector spin-wave energy at low temperatures with neutron scattering, it was found that the out-of-plane FM in the 2D honeycomb lattice CrI$_3$ is controlled by spin-orbit coupling (SOC) induced magnetic anisotropy, instead of magnetic exchange coupling as in a conventional ferromagnet.[4] Moreover, the SOC induced magnetic anisotropy can be tuned by substituting some chloride atoms with bromine ones in the similar chromium halide $CrCl_{3-x}Br_x$ system, achieving a magnetization pointing toward any spatial direction, while the easy-axis of $CrCl_3$ is in-plane and the one of $CrBr_3$ is out-of-plane.[5] This demonstrates the tuning of magnetization in chromium halides with mixed halide chemistry and can provide more flexibility in design for compact 2D magnetic devices.

---

1 Email: hongtang@temple.edu,   ORCiD: 0000-0002-6221-0922

Huang et al.[1] studied the magneto-optical (MO) Kerr effect with an applied magnetic field perpendicular to the 2D plane for $CrI_3$ monolayer and multilayer and revealed the large Kerr rotation angles $\theta_K$ of about 10 mrad for a monolayer, and about 70 mrad for a trilayer, and the metamagnetic effect for a bilayer. Seyler et al.[6] revealed the spontaneous circularly polarized photoluminescence in monolayer $CrI_3$ under linearly polarized light excitation, with a magnetization direction dependent helicity. Electric fields can also induce magnetic effects in $CrI_3$ layers. Both Jiang et al.[7] and Huang et al.[8] observed a large linear magnetoelectric (ME) effect in the $CrI_3$ bilayer field-effect device, in which an externally applied out-of-plane electric field can reversibly switch the bilayer between FM and AFM (antiferromagnetic) states. Moreover, with $CrI_3$-graphene vertical heterostructures, Jiang et al.[9] enabled the electron or hole doping by the gate voltage and demonstrated the tunability of magnetic ordering strength in monolayer $CrI_3$, the strong doping-dependent interlayer exchange coupling and the robust switching of magnetization in bilayer $CrI_3$ by the gate voltage. Those findings suggest the unique potential of the 2D magnets for atomically thin magnetic and optical nanodevices, electrically controlled spintronic and valleytronic applications through proximity coupling in van der Waals homo- or heterostructures.

The optical absorption and magneto-optical properties of monolayers $CrI_3$ and $CrBr_3$ were extensively investigated by Wu et al.[10,11] with the developed SOC enabled spinor GW[12,13] and BSE[14] many-body perturbation approach, which allows the accurate calculation of spin-splitting, self-energy corrections, electron-hole interactions, and the diagonal and off-diagonal elements of the layer's macroscopic dielectric tensor, and hence reasonably reproduces the experimentally measured optical absorption and MO Faraday angles for monolayer $CrI_3$. It showed that the optical and MO responses of 2D magnetic semiconductors are strongly enhanced by the SOC, self-energy, and excitonic effects.

Nanoribbon is a practical form for applications of 2D materials in nanodevices[15-17], since strips or flakes of 2D layers are usually cut off and placed in the devices, especially in the spintronic devices.[15] The further reduced dimensionality of nanoribbons usually leads to enhanced quantum confinements and provides unique edge states, which enable more functionality.[15] Both Jiang et al.[18] and Wang et al.[19] calculated the magnetic properties of $CrI_3$ nanoribbons and confirmed that FM is the stable state for $CrI_3$ nanoribbons, and the edge states dominate the band structures around the Fermi level. By applying uniaxial strains or bending, one can create local strains on the nanoribbons and effectively modify the edge states and band structures[20-23], leading to controllable electronic and optical properties. In this work, we investigate tensile strains and bending induced tunability of electronic structures of the magnetic $CrI_3$ nanoribbons. Our DFT calculations[24-28] show that for narrow $CrI_3$ nanoribbons, the ferromagnetic state with out-of-plane spin moments is almost equally stable with the one with in-plane spin moments, different from the monolayer $CrI_3$. The recently developed mTASK metaGGA[28] gives improved magnetic moments and edge band gaps for $CrI_3$ nanoribbons, showing an improved description for p-orbital derived edge states. We also use the full spinor enabled GW-BSE to study the optical absorption and magneto-optical properties of the $CrI_3$ magnetic nanoribbon. We found that there are many dark exciton states with low energy in the $CrI_3$ nanoribbon, and the dark excitons are not Frenkel-like excitons, different from the monolayer $CrI_3$ case. The exciton states, optical absorption and magneto-optical property all show manifest tunability with bending or strains. This suggests the application of the magnetic nanoribbon in controllable magnetic optoelectronics, 1D or 2D magnetic storage nanodevices, and spin-based quantum information processes.

**RESULTS AND DISCUSSION**

**Band structures and gaps.** We build our nanoribbons from the pre-relaxed $CrI_3$ monolayer. Our PBE-SOC relaxed $CrI_3$ monolayer has the 2D hexagonal lattice constant $a = b = 6.99$ Å, which is close to the 6.98 Å reported in Ref. 10. The nanoribbons are cut from the monolayer $CrI_3$ with iodine (I) atom terminated edges. The structures of the nanoribbons are in Figures 1(a)-(e). The Cr atoms close to the edges form zigzag chains along the ribbon periodical direction, which is the supercell vector $c$ (or axis $z$). The nanoribbon supercells with 4, 6, and 8 Cr atoms are studied and denoted as $Z4CrI_3$, $Z6CrI_3$ and $Z8CrI_3$, respectively. The relaxed structures of nanoribbons under different bending curvatures $\kappa$ ($\kappa = 1/R$, R is the bending curvature radius) or tensile strains $\epsilon$ are shown in Figures 1(f)-(h). The supercell vectors $a$, $b$ and $c$ are orientated along axes $x$, $y$ and $z$, respectively. The relaxed structures of narrow $Z4CrI_3$ (with ribbon width ~12.4 Å) are almost flat. Therefore, bending produces an effective compression along the ribbon's width direction. $Z6CrI_3$ (width ~18.5 Å) and $Z8CrI_3$ (width ~24.5 Å) nanoribbons show relatively large deformations around the ribbon's middle region, while the regions close to the edges are relatively flat. The calculated band structures are shown in Figures 1(i)-(l). After relaxations, the length of vector $c$ of the supercell (LC) (Figure 1(m)), which measures the lattice constant along the periodic direction of nanoribbons, slightly increases from a larger tensile strain towards flat, and to larger bending curvatures. $Z6CrI_3$ has a larger LC than $Z4CrI_3$ and $Z8CrI_3$ at larger bending curvatures, due to its two asymmetric edges about the middle of the ribbon (Figure 1(d)). The band structures (Figures 1(i)-(l)) show almost flat conduction bands near the Fermi level. The partial density of states (pDOS) analysis (Figure 1(n)) shows a large contribution to the valence band edge (close to the Fermi level) from the p-orbital of both the edge and the non-edge I atoms, while the nearly degenerate two flat edge conduction bands right above the Fermi level are dominated by the p-orbital of the edge I atoms. The lower valence bands (energy ~0.5-1.0 eV) have both contributions from the p-orbital of I and d-orbital of Cr atoms. The conduction band continuum (~1.0-1.5 eV) above the edge bands is derived from the Cr d-orbitals and I p-orbitals with minor contribution from the edge I atoms, while the higher conduction bands (over 2.5 eV) are mainly derived from the Cr d-orbitals. mTASK[28] gives band structures of $Z4CrI_3$ similar to those from PBE-U[27] around the Fermi level, while it pushes up the higher conduction bands (which begin about 2.5 eV above the Fermi level in PBE-U) derived mainly from the Cr d-orbitals by about 1.5 eV more than PBE-U does. The change of band structures of wider nanoribbons with bending is more obvious, as shown in Figures 1(k) and 1(l) for $Z6CrI_3$ and $Z8CrI_3$.

The band gaps are shown in Figures 2(a)-(c). While $Z4CrI_3$ shows a slight variation of gaps with bending and tensile strains, LSDA-U[27], mTASK, and $G_0W_0$[12] all show the highest gap at $\kappa = 0.125/$Å (or R = 8 Å). For the edge gaps (EG), as seen in Figure 2(a), mTASK gives higher values among the density functional approximations used here and is closer to the many-body perturbation method $G_0W_0$ evaluated with LSDA-U reference. For non-edge gaps (NEG), as seen in Figure 2(b), SCAN[25] gives larger values than other DFTs, however, with an oscillating trend in the bending curvature region, while PBE,[24] PBE-U,[27] mTASK, and LSDA-U all give similar values for NEG. $r^2SCAN$[29] results are similar to SCAN results for NEG and are slightly lower than those of SCAN for NEG. $G_0W_0$ shows large values for both EG (~2.3 eV) and NEG (~3.1 eV), indicating the large self-energy correction to the quasiparticle energy for the nanoribbon systems. TASK, mTASK, SCAN and $r^2SCAN$ are meta-GGA density functionals with ingredients related to the electron kinetic energy density, with improved description of band gaps. The p-orbital EG of $Z4CrI_3$ is better described by mTASK, consistent with reports for other p-orbital materials.[28,30] The change of EG with bending is larger for wider nanoribbons $Z6CrI_3$ and $Z8CrI_3$, as shown in Figure 2(c).

**Magnetic moments.** The average magnetic moment of the four Cr atoms in the cell of Z4CrI$_3$ nanoribbon with different tensile strains and bending curvatures is shown in Figure 2(d). Table 1 shows the calculated magnetic moments on atoms in the Z4CrI$_3$ flat nanoribbon. All DFT results in Table 1 show approximately the same magnetic moment of 0.21-0.26 $\mu_B$ of the edge I atoms (atom indices 10-13), while LSDA-U, PBE, SCAN and r$^2$SCAN give similar values (~3.0-3.2 $\mu_B$) for Cr atoms, see Figure 2(e) for atom indexes, and both mTASK and TASK give the values (~3.5-3.6 $\mu_B$) close to those of PBE-U (~3.3 $\mu_B$). The mTASK value is very close to the DMC (diffusion Monte Carlo) value[31] (3.62 $\mu_B$) for Cr in monolayer CrI$_3$. All DFT results here give the moments on Cr and edge I atoms pointed to the positive y direction and perpendicular to the ribbon plane, while the relatively small moments on other non-edge I atoms point to the negative y direction, indicating that those I atoms have a weak antiferromagnetic coupling with Cr and edge I atoms). It was suggested that Cr atoms are coupled by super-exchange via p-orbitals of the neighboring I atoms. The edge I atoms only connect with one Cr atom and have unsaturated dangling bonds, and hence ferromagnetically couple with Cr atoms.

Table 2 shows the energy difference between cases with different directions of the magnetic field for the Z4CrI$_3$ nanoribbon under different tensile and bending, calculated with PBE-U and the SOC effect included. The direction of magnetic field is selected by the orientation of the spin quantization axis. Orientation 010 represents the direction of vector b (or axis y) and orientation 001 the direction of vector c (or the periodical z axis), and so on. As can be seen, the three orientations (011, 111 and 101) surrounding the 001 orientation have higher energies than the 001 orientation, and similarly, the three orientations (011,111 and 110) surrounding the 010 orientation also have higher energies than the orientation 010. However, orientation 100 has higher energy than its three surrounding orientations (110, 111 and 101). This means the two orientations 001 (in-plane along the z direction) and 010 (out-of-plane) are locally stable and they are more favorable directions of magnetic moments than other directions. This is different from the monolayer CrI$_3$ case, where the out-of-plane configuration is more stable. We believe that this effect is due to the in-plane rotational symmetry breaking by the ribbon edges and the interaction between edge states. Unlike the monolayer case, which only has two stable magnetization directions (up and down) perpendicular to the plane, a CrI$_3$ nanoribbon can have four possible ones; up, down, forward (along the ribbon direction) and backward magnetization directions. This feature can be used for magnetic nanodevices with tunable switching of magnetization direction. The average magnetic moment of Cr atoms in Z4CrI$_3$ shows little variations with bending curvatures (Figure 2(d)), except for the results of SCAN, where a similar vibrating trend is shown in line with that of NEG. This may relate to the numerical stability of SCAN previously reported.[29] r$^2$SCAN is based on the SCAN framework but improves the smoothness of the interpolation function and hence the numerical stability.[29]

The similar trend is also shown for the Z6CrI$_3$ nanoribbon, as seen in Table 3, where the two orientations 001 and 010 are locally stable, like in the case of the Z4CrI$_3$ nanoribbon. The situation is slightly different for the Z8CrI$_3$ nanoribbon, as shown in Table 4. It shows that orientation 001 is locally stable for all the bending curvatures. For orientation 010, from flat to R11, the three orientations (011,111 and 110) surrounding the 010 orientation have higher energies, however, for R7 and R6, orientation 011 has lower energies than the 010 orientation. This means that for small bending curvatures, orientation 010 is still a locally stable direction, however, for larger bending curvatures, orientation 010 is no longer locally stable, and only orientation 001 is the stable one. The slightly different trend may relate to the wider width of the Z8CrI$_3$ nanoribbon, in which large bending curvatures cause a large deformation and structure alteration around the middle region of the ribbon and can alter the SOC strength accordingly, making the magnetization parallel to the ribbon axis more favorable.

Figure 3 shows the magnetic moments on Cr atoms in the Z4CrI$_3$, Z6CrI$_3$ and Z8CrI$_3$ nanoribbons as a function of the bending or tensile strains for two magnetic moment directions, namely in-plane and out-of-plane. For the in-plane magnetic moment orientation, as shown in Figure 3(a), for the Z4CrI$_3$ nanoribbon, with a decrease in tensile strain and an increase in bending, the magnetic moments on Cr atoms are generally decreasing, especially for the Cr atom in middle region of the ribbon with a maximum decrease of about 1.2%. Since the width of the Z4CrI$_3$ nanoribbon is relatively small (width ~12.4 Å), the relaxed structures of Z4CrI$_3$ are almost flat, see Figure 1(f). Therefore, bending produces an effective compression along the ribbon's width direction. This makes the magnetic moments of Cr atoms generally decrease. For the Z6CrI$_3$ nanoribbon, as in Figure 3(b), the relaxed structure with a small bending (R = 10Å) does not curve or bend so much and is close to a flat one. This also produces an effective compression along the width direction of the nanoribbon and, like in the case of the Z4CrI$_3$ nanoribbon, makes the magnetic moments on Cr atoms decrease. However, with larger bending curvatures (R < 10Å), the relaxed structures of the Z6CrI$_3$ nanoribbon are substantially curved. This makes the magnetic moments on Cr atoms generally increase with a further increase in bending curvatures. The similar trend is shown for the even wider Z8CrI$_3$ nanoribbon, as shown in Figure 3(c), where for bending curvature radii R = 16, 11, 7, 6 Å, the relaxed structures of bent nanoribbons are all substantially curved, and this results in an increasing trend of the magnetic moments on Cr atoms with bending curvatures. The situation is similar for the out-of-plane magnetic moment direction, as shown in Figures 3(d)-(f).

Based on a 1D Ising spin chain model,[32] we calculate the spin correlation length in the flat Z4CrI$_3$ nanoribbon and its temperature dependence, see note 1 and Figure S1 in the supplemental information. The plots of the calculated spin correlation length $\xi(T)$ for flat Z4CrI$_3$ nanoribbon are shown in Figure S1. The curve shows a steep increase below 25 Kelvin. Around 27 K, the spin correlation length is still about 10 Å, as seen in Figure S1(b). Thus, we estimate the ferromagnetic transition temperature $T_c = 27$ K, which is smaller than the measured $T_c = 45$ K for the monolayer CrI$_3$.[1] However, with a reduced ferromagnetic transition temperature, CrI$_3$ nanoribbons still show stable magnetic properties at low temperatures. The zigzag CrI$_3$ nanoribbons that we calculated have iodine atom terminated edges. The relative stability of those CrI$_3$ nanoribbons were studied previously.[19] This study shows that the edge iodine atoms terminate the dangling bonds in the near-edge Cr atoms, and the energy cost of having the edge Cr atoms with dangling bonds is higher than that of having edge I atoms. In addition, zigzag CrI$_3$ nanoribbons are stabler than armchair ones, due to dimerization of the zigzag CrI$_3$ I atoms. The hydrogen passivation to the zigzag CrI$_3$ nanoribbons will less drastically change the overall magnetic ordering and magnetization strength, since the edge I atoms have much lower magnetic moments (< 0.3 $\mu_B$), compared with that of Cr atoms (~3.0-3.5 $\mu_B$), see Table 1. The contribution of edge I atoms to the density of states (DOS) in the occupied valence bands is small, see Figure 1(n). The CrI$_3$ nanoribbons studied here can be made in a high vacuum environment by the electron beam etching or scanning tunnelling lithography techniques[17] and capsulated with wide bandgap nanosheets, such as hBN. This will preserve the edge condition of the crafted nanoribbons.

**Optical absorptions and exciton states.** The calculated absorption spectra of Z4CrI$_3$ are shown in Figures 4(a)-(e). The spin resolved GW quasiparticle band structures under different tensile strains and bending are shown in Figures 4(f)-(k). The three main peaks A, B, C below the quasiparticle gap indicate the strong excitonic effect in nanoribbons. From flat to R8, the three peaks move slightly to higher energies, maybe due to the slightly higher EG and NEG gaps of R8. From R8 to R5, both EG and NEG decrease, the three peaks move to low energies by ~0.1-0.2 eV. The peak positions barely change from flat to T2, due to a relatively small strain. Further increasing the tensile strain to 5%, the peaks move towards low energies. The height of peak B is more affected by the tensile strain and bending, showing an increase with tensile strains and a decrease with bending curvatures. In all tensile and bending cases, there are several dark excitons within an energy range 0.5-1.1 eV lower than that of peak A (Supplemental Figure S2). Those dark excitons are mainly contributed by the transitions from valence bands V1 to V11 to the edge conduction band C1 and C2, see Figures 5(a) and (b) for plots of dark excitons at 0.94 and 0.97 eV. Figures 5(a)-(f) show the wavefunctions both in real and reciprocal spaces and the electron-band and hole-band composition of several typical excitons of the Z4CrI$_3$ flat nanoribbon. The edge C1/C2 are mainly from the p-orbitals of edge I atoms with a minority spin polarization, as seen in Figures 4(f)-(k), while the valences V1-V11 have a mixed spin polarization. By analyzing the spin polarization of the involved valence and conduction states of the excitons, we plot the spin polarization resolved valence (or hole) and conduction (or electron) states involved in excitons in the Z4CrI$_3$ nanoribbon, as shown in Figures 6(a)-(i). It shows that for dark excitons (Figures 6(a)-(d)) the electron states are of minority spin and most of the hole states are of majority spin, indicating the formation of the dark excitons from electrons and holes of unlike spin polarizations, consistent with that of the dark excitons in the CrI$_3$ monolayer[10]. However, due to the presence of the edges, the dark excitons in nanoribbons have a more extended region in real space and can span the ribbon width (subplots I-III in Figures 5(a) and (b)), indicating non-Frenkel-like excitons, different from the Frenkel-like dark exciton in monolayer CrI$_3$.

Due to strong SOC effects, the Bloch wave functions of the electron and hole states composing an exciton in the system have various spin polarization directions. This may give rich and complex excitonic spin configurations. Peak A (peak B) is composed of excitons with transitions mainly from V1-V5 to C1/C2 (from V1-V7 to C1/C2), see Figures 5(c) and (d) for the band transition decomposition of the bright excitons at 1.25 eV in peak A and at 1.58 eV in peak B, which have the highest oscillator strength in the peaks. The valence bands V3-V7 have mixed spin polarizations with both majority and minority ones, therefore, the spin configuration of excitons forming peaks A and B is generally complex. The exciton states with the highest oscillator strength in peaks A and B have contributions of transitions from both majority and minority spin valence states to the minority spin conduction edge states (Figures 6(e) and (f)), indicating a rich and mixed spin configuration for those bright excitons, while in a CrI$_3$ monolayer, the bright bound excitons are mainly of like-spin polarization of electron and hole states.[10] The excitons of the highest oscillator strength in peak C and the smaller peak D all have a diverse contribution of transitions from V1-V11 to C1/C8, see Figures 5(e) and (f) for the band decomposition of excitons at 1.85 eV and 2.03 eV. This also leads to the formation of bright excitons from electrons and holes of mixed spin polarizations. The excitons at 1.25 eV in peak A and at 1.58 eV in peak B have a more extended spatial region along the two edges, with the former mainly localized around Γ in the momentum (k) space and the later nodal in k space and slightly nodal in real space, while the excitons at 1.85 and 2.03 eV in peaks C and D are slightly nodal in the k space and very nodal in the real space with an obvious concentration of electron wavefunction around the vicinity of the hole, mainly due to the relatively wide range of conduction and valence bands involved in the formation of those excitons. As can be seen in Figures 6(g) and (h), the exciton at 1.85 eV also has a mixed spin configuration and the spin polarizations of hole and electron involved in the exciton can be a superposition of majority and minority spins.

Exciton lifetimes, and in general exciton dynamics, have a critical role in the mechanism of photocatalyst materials such as organic[33] and inorganic quantum dots,[34] nanoplatelets,[34] and monolayers.[35,36] We also calculated the intrinsic lifetimes[37] of excitons in the Z4CrI$_3$ nanoribbon under different tensile strains and bending (Supplemental Figures S4). They show an obvious dependence on the strains and bending, and some excitons within the regime of 0.5~2.0 eV show a lifetime of several nanoseconds or higher. We also think that these features can be suitable for exploring the application of CrI$_3$ nanoribbons in exciton-based quantum information processes.

**Magneto-optical properties.** The zigzag Z4CrI$_3$ nanoribbon has a mirror symmetry about the plane (parallel to the yz-plane) through the middle of the ribbon. It leads to two elliptically polarized modes (EPM) propagating along the x direction. To demonstrate its MO properties, we calculated the magnetic dichroism (MDC), Faraday and MO Kerr effects for the Z4CrI$_3$ nanoribbon. The results are shown in Figures 7(a)-(f) for the flat nanoribbon. Figure 7(a) shows the very different absorption spectra of the two opposite (left and right) EPMs, denoted as $\sigma^+$ and $\sigma^-$. This leads to a large MDC contrast $\eta$, and $\eta = (\text{Abs}(\sigma^+) - \text{Abs}(\sigma^-))/(\text{Abs}(\sigma^+) + \text{Abs}(\sigma^-))$, where $\text{Abs}(\sigma^\pm)$ denotes the absorbance of $\sigma^\pm$, as shown in Figure 7(b), especially within the region of 1.0-2.0 eV, manifesting the significant effect of excitons to MDC. For the transmitted Faraday effect, the rotation angle $\theta_F$ and ellipticity $\chi_F$ (see Ref. 10 for definitions) can be about 2~3 mrad within the region of 1.0-2.0 eV with the GW-BSE results (Figure 7(c)), while the results without electron-hole (e-h) interactions (GW-RPA) (Figure 7(d)) show a zero ellipticity $\chi_F$ and a slightly increasing rotation angle $\theta_F$. This demonstrates the inclusion of e-h interaction is important. The slightly increasing $\theta_F$ from GW-RPA may be due to the normal optical anisotropy along the $y$ and $z$ directions. Similarly, for the reflective MO Kerr effect, the rotation $\theta_K$ and ellipticity $\chi_K$ can be as large as 600~700 mrad within the region of 1.0-2.0 eV from GW-BSE (Figure 7(e)), indicating a strong MO Kerr effect. The results without e-h interaction (Figure 7(f)) give a zero rotation $\theta_K$ and an almost constant ~600 mrad ellipticity $\chi_K$, also due to the normal optical anisotropy along axes y and z. Bending and tensile strains can change the MDC, Faraday and MO Kerr signals manifestly, as shown in Figures 7(g)-(k). For example, within the region of 1.0 -1.2 eV, the region of mainly peak A, the case of bending R5 has a large difference from other ones. This demonstrates the tunability of those properties with bending and tensile strains, which can be harnessed for controllable optoelectronic devices.

## CONCLUSION

In summary, from first principles, we have calculated the band structures, band gaps, and magnetic moments of zigzag CrI$_3$ nanoribbons under different tensile strains and bending curvatures with density functional approximations, and their optical absorption spectra with the many-body perturbation GW-BSE method, and investigated their magneto-optical properties, including magnetic dichroism, Faraday and magneto-optical Kerr effects. The flat conduction bands close to the Fermi level are dominated by the p-orbital of the edge iodine atoms, while the valence bands close to the Fermi level are mostly from the p-orbital of I atoms (both edge and non-edge) and to a lesser extent from Cr d-orbitals. The conduction band beyond the edge bands mostly consists of Cr d-orbitals. The edge I atoms ferromagnetically couple with Cr atoms, while the non-edge I atoms weakly couple antiferromagnetically. Unlike the more stable out-of-plane magnetic moments of a CrI$_3$ monolayer, narrow CrI$_3$ nanoribbons (Z4CrI$_3$) have two locally stable magnetic moment directions, namely out-of-plane and in-plane (along the ribbon axis) with an energy difference ~0.3 meV/atom between the two directions. Based on the one-dimensional Ising spin chain model, the spin correlation length of the narrow CrI$_3$ nanoribbon is estimated as about 10 Å at its estimated Curie temperature of 27 Kelvin, which it lower than the measured 45 Kelvin of the monolayer CrI$_3$. The band structures and orbital characters are vital for the formations of the dark and bright bound excitons in the narrow CrI$_3$ nanoribbons. The dark excitons with energies lower than that of the lowest-energy absorption peak are mainly formed from holes and edge electrons with unlike spin polarizations and are non-Frenkel-like excitons, while the bright excitons in the main absorption peaks are from holes and electrons with mixed spin polarizations. Tensile strains and bending can manifestly modulate the absorption spectrum, magnetic dichroism, Faraday and magneto-optical Kerr effects of the CrI$_3$ nanoribbons within a technologically important photo energy range of ~1.0-2.0 eV, suggesting a potential application in tunable magnetic optoelectronic nanodevices. It is worthy of noting that some nanoribbons, such as zigzag $MoS_2$ [38] and graphene nanoribbons,[17] show magnetic ordering on ribbon edges, due to edge structure effects. For example, the graphene zigzag nanoribbons (GZNRs)[17] have a magnetic property dominated by the carbon atoms on the two edges, which belong to two different sublattices in the honeycomb lattice, while the contributions from the atoms within the inner ribbon region is small. Narrow GZNRs are predicted and experimentally confirmed as semiconductive and the two edges are coupled antiferromagnetically, while wide GZNRs (width > 8 nm) are metallic, and the two edges are coupled ferromagnetically. Within each edge of GZNRs, the edge carbon atoms are coupled ferromagnetically, with the magnetic moment mainly perpendicular to the ribbon plane. The ferromagnetism in $CrI_3$ nanoribbons stems from its parent 2D form, in which all Cr atoms contribute to the total magnetization, and it is robust. The magnetization strength in $CrI_3$ nanoribbons is less influenced by edge effects. The $CrI_3$ nanoribbons can be used in 1D or 2D magnetic storage nanodevices, tunable compact magnetic optoelectronics, and spin-based quantum information controls.

## METHODS

**Computational details.** Density functional theory (DFT) calculations were conducted in the Vienna Ab initio Software Package (VASP)[39] with projector augmented-wave pseudopotentials.[40] PBE,[24] SCAN,[25] TASK[26] and mTASK,[28] approximations with spin-orbit coupling (SOC) were used to calculate the band structures of nanoribbons. In the PBE-U calculation, a rotationally invariant approach with U=2.65 eV[41] for Cr atoms was used. The vacuum layer of more than 15 Å is added along the direction of nanoribbon width and along the out-of-plane direction of the nanoribbon, to avoid the interactions with periodic images. The energy cutoff is 500 eV. The k-point mesh of $1 \times 1 \times 7$ was used for relaxation and $1 \times 1 \times 24$ was used for electronic calculations. All nanoribbons were fully structurally relaxed with PBE with all forces less than 0.02 eV/Å. During the relaxation, the x and y coordinates of the two outer most metal atoms close to the two edges were fixed, while their coordinates along the ribbon axis direction, and all the coordinates of other atoms were allowed to relax. The spinor $G_0W_0$ and $G_0W_0$+BSE calculations were conducted in BerkeleyGW[12] by pairing with Quantum ESPRESSO.[42] The $G_0W_0$ is based on the LSDA-U calculations with U = 1.5 eV and J = 0.5 eV for Cr atoms.[10] The wavefunction energy cutoff is 70 Ry (~950 eV). The energy cutoff for the epsilon matrix is 14 Ry (~190 eV). The k-point mesh of $1 \times 1 \times 64$ (interpolated from $1 \times 1 \times 16$) and the valence bands of 15 and conduction bands of 10 were set for optical absorption calculations. The band number for summation is 800. The correction of the exact static remainder and the wire Coulomb truncation for 1D systems were also used.

## Supporting Information

Technical Note 1 for calculating the spin correlation length with the 1D Ising spin chain model, note 2 for the optical eigenmodes in the CrI3 nanoribbons, Figure S1 for the calculated spin correlation length, Figure S2 for the exciton energy spectra, Figure S3 for the crystal structure of the Z4CrI₄ nanoribbon, and Figure S5 for the calculated exciton lifetimes.

## Author Contributions

H.T. and A.R. designed the research; H.T. conducted the computation and analyzed the data with S.N.; H.T. wrote the manuscript with discussions with S.N., Q.Y. and A.R.; all authors discussed the results and contributed to the manuscript.

## Notes

The authors declare no competing financial interest. The data that support the findings of this study are available from the corresponding author upon reasonable request.

## Acknowledgement

This material is based upon work supported by the U.S. Department of Energy, Office of Science, Office of Basic Energy Sciences, under Award Number DE-SC0021263. This research used resources of the National Energy Research Scientific Computing Center, a DOE Office of Science User Facility supported by the Office of Science of the U.S. Department of Energy under Contract No. DE-AC02-05CH11231.

## References

1.  Huang, B.; Clark, G.; Navarro-Moratalla, E.; Klein, D.; Cheng, R.; Seyler, K. L.; Zhong, D.; Schmidgall, E.; McGuire, M. A.; Cobden, D. H.; Yao, W.; Xiao, D.; Jarillo-Herrero, P.; Xu, X. Layer-dependent ferromagnetism in a van der Waals crystal down to the monolayer limit. *Nature* **2017**, *546*, 270–273.

2.  Mermin, N. D.; Wagner, H. Absence of ferromagnetism or antiferromagnetism in one- or two-dimensional isotropic Heisenberg models. *Phys. Rev. Lett.* **1966**, *17*, 1133–1136.

3.  Hutchings, M. T.; Als-Nielsen, J.; Lindgard, P. A.; Walker, P. J. Neutron scattering investigation of the temperature dependence of long-wavelength spin waves in ferromagnetic $Rb_2CrCl_4$. *J. Phys. C: Solid State Phys.* **1981**, *14*, 5327.

4.  Chen, L.; Chung, J.-H.; Chen, T.; Duan, C.; Schneidewind, A.; Radelytskyi, I.; Voneshen, D. J.; Ewings, R. A.; Stone, M. B.; Kolesnikov, A. I.; Winn, B.; Chi, S.; Mole, R. A.; Yu, D. H.; Gao, B.; Dai, P. Magnetic anisotropy in ferromagnetic $CrI_3$. *Phys. Rev. B* **2020**, *101*, 134418.

5.  Abramchuk, M.; Jaszewski, S.; Metz, K. R.; Osterhoudt, G. B.; Wang, Y.; Burch, K. S.; Tafti, F. Controlling Magnetic and Optical Properties of the van der Waals Crystal $CrCl_{3-x}Br_x$ via Mixed Halide Chemistry. *Advanced materials* **2018**, *30*, 1801325.

6.    Seyler, K. L.; Zhong, D.; Klein, D. R.; Gao, S.; Zhang, X.; Huang, B.; Navarro-Moratalla, E.; Yang, L.; Cobden, D. H.; McGuire, M. A.; Yao, W.; Xiao, D.; Jarillo-Herrero, P.; Xu, X. Ligand-field helical luminescence in a 2D ferromagnetic insulator. *Nature Physics* **2018**, *14*, 277–281.

7.    Jiang, S.; Shan, J; Mak, K. F. Electric-field switching of two-dimensional van der Waals magnets. *Nature Materials* **2018**, *17*, 406–410.

8.    Huang, B.; Clark, G.; Klein, D. R.; MacNeill, D.; Navarro-Moratalla, E.; Seyler, K. L.; Wilson, N.; McGuire, M. A.; Cobden, D. H.; Xiao, D.; Yao, W.; Jarillo-Herrero, P.; Xu, X. Electrical control of 2D magnetism in bilayer $CrI_3$. *Nature Nanotechnology* **2018**, *13*, 544–548.

9.    Jiang, S.; Li, L.; Wang, Z.; Mak, K. F.; Shan, J. Controlling magnetism in 2D $CrI_3$ by electrostatic doping. *Nature Nanotechnology* **2018**, *13*, 549–553.

10.   Wu, M.; Li, Z.; Cao, T.; Louie, S. G. Physical origin of giant excitonic and magneto-optical responses in two-dimensional ferromagnetic insulators. *Nature Commun.* **2019**, *10*, 2371.

11.   Wu, M.; Li, Z.; Louie, S. G. Optical and magneto-optical properties of ferromagnetic monolayer $CrBr_3$: A first-principles GW and GW plus Bethe-Salpeter equation study. *Phys. Rev. Materials* **2022**, *6*, 014008.

12.   Deslippe, J.; Samsonidze, G.; Strubbe, D. A.; Jain, M.; Cohen, M. L.; Louie, S. G. BerkeleyGW: A Massively Parallel Computer Package for the Calculation of the Quasiparticle and Optical Properties of Materials and Nanostructures. *Comput. Phys. Commun.* **2012**, *183*, 1269.

13.   Hybertsen, M. S.; Louie, S. G. Electron correlation in semiconductors and insulators: Band gaps and quasiparticle energies. *Phys. Rev. B* **1986**, 34, 5390.

14.   Rohlfing, M.; Louie, S. G. Electron-hole excitations and optical spectra from first principles. *Phys. Rev. B* **2000**, *62*, 4927.

15.   Wang, H.; Wang, H. S.; Ma, C.; Chen, L.; Jiang, C.; Chen, C.; Xie, X.; Li, A.-P.; Wang, X. Graphene nanoribbons for quantum electronics. *Nature Reviews Physics* **2021**, *3*, 791-802.

16.   Lou, P. Quasiparticle Energies, Exciton Level Structures and Optical Absorption Spectra of Ultra-Narrow ZSiCNRs. *RSC Adv.* **2017**, *7*, 52053-52064.

17.   Magda, G. Z.; Jin, X.; Hagymási, I.; Vancsó P.; Osváth Z.; Nemes-Incze, P.; Hwang, C.; Biró, L. P.; Tapasztó, L. Room-temperature magnetic order on zigzag edges of narrow graphene nanoribbons. *Nature*, **2014**, *514*, 608-611.

18.   Jiang, W.; Li, S.; Liu, H.; Lu, G.; Zheng, F.; Zhang, P. First-principles calculations of magnetic edge states in zigzag $CrI_3$ nanoribbons. *Physics Letters A* **2019**, *383*, 754-758.

19.   Wang, J.-Z.; Huang, J.-Q.; Wang, Y.-N.; Yang, T.; Zhang, Z.-D. Electronic and magnetic properties of $CrI_3$ nanoribbons and nanotubes. *Chin. Phys. B* **2019**, *28*, 077301.

20.   Neupane, B.; Tang, H.; Nepal, N. K.; Ruzsinszky, A. Bending as a control knob for the electronic and optical properties of phosphorene nanoribbons. *Phys. Rev. Materials* **2022**, *6*, 014010.

21.   Tang, H.; Neupane, B.; Neupane, S.; Ruan, S.; Nepal, N. K.; Ruzsinszky, A. Tunable band gaps and optical absorption properties of bent MoS2 nanoribbons. *Scientific Reports* **2022**, *12*, 3008.

22.    Yu, L.; Ruzsinszky, A.; Perdew, J. P. Bending two-dimensional materials to control charge localization and Fermi-level shift. *Nano Lett*. **2016**, *16*, 2444-2449.

23.    Nepal, N. K.; Yu, L.; Yan, Q.; Ruzsinszky, A. First-principles study of mechanical and electronic properties of bent monolayer transition metal dichalcogenides. *Phys. Rev. Materials*. **2019**, *3*, 073601.

24.    Perdew, J. P.; Burke, K.; Ernzerhof, M. Generalized Gradient Approximation Made Simple. *Phys. Rev. Lett.* **1996**, *77*, 3865.

25.    Sun, J.; Ruzsinszky, A.; Perdew, J. P. Strongly constrained and appropriately normed semilocal density functional. *Phys. Rev. Lett*. **2015**, *115*, 036402.

26.    Aschebrock, T.; Kümmel, S. Ultranonlocality and accurate band gaps from a meta-generalized gradient approximation. *Phy. Rev. Res*. **2019**, *1*, 033082.

27.    Cococcioni, M.; de Gironcoli, S. Linear response approach to the calculation of the effective interaction parameters in the LDA+U method. *Phys. Rev. B* **2005**, *71*, 035105.

28.    Neupane, B.; Tang, H.; Nepal, N. K.; Adhikari, S.; Ruzsinszky, A. Opening band gaps of low-dimensional materials at the meta-GGA level of density functional approximations. *Phys. Rev. Materials* **2021**, *5*, 063803.

29.    Furness, J. W.; Kaplan, A. D.; Ning, J.; Perdew, J. P.; Sun, J. Accurate and numerically efficient r2SCAN meta-generalized gradient approximation. *J. Phys. Chem. Lett*. **2020**, *11*, 8208−8215.

30.    Tran, F.; Doumont, J.; Kalantari, L.; Blaha, P.; Rauch, T.; Borlido, P.; Botti, S.; Marques, M. A.; Patra, A.; Jana, S.; Samal, P. Bandgap of two-dimensional materials: Thorough assessment of modern exchange–correlation functionals. *J. Chem. Phys*. **2021**, *155*, 104103.

31.    Staros, D.; Hu, G.; Tiihonen, J.; Nanguneri, R.; Krogel, J.; Chandler Bennett, M.; Heinonen, O.; Ganesh, P.; Rubenstein, B. A combined first principles study of the structural, magnetic, and phonon properties of monolayer CrI$_3$. *J. Chem. Phys*. **2022**, *156*, 014707.

32.    Ashcroft, N. W.; Mermin, N. D. *Solid State Physics*; Thomson Press (India) Ltd, 2003, ISBN-10: 8131500527.

33.    Mikhnenko, O. V.; Blom, P. W. M.; Nguyen, T. -Q. Exciton diffusion in organic semiconductors. *Energy Environ. Sci*. **2015**, *8*, 1867.

34.    Li, Q.; Lian, T. Exciton dissociation dynamics and light-driven H2 generation in colloidal 2D cadmium chalcogenide nanoplatelet heterostructures. *Nano Res*. **2018**, *11*, 3031.

35.    Giebink, N. C.; Wiederrecht, G. P; Wasielewski, M. R.; Forrest, S. R. Thermodynamic efficiency limit of excitonic solar cells. *Phys. Rev. B* **2011**, *83*, 195326.

36.    Wang, H.; Zhang, X.; Xie, Y. Recent progress in ultrathin two-dimensional semiconductors for photocatalysis. *Mater. Sci. Eng. R* **2018**, *130*, 1.

37.    Spataru, D. C.; Ismail-Beigi, S.; Capaz, R. B.; Louie, S. G. Theory and Ab Initio Calculation of Radiative Lifetime of Excitons in Semiconducting Carbon Nanotubes. *Phys. Rev. Lett*. **2005**, *95*, 247402.

38.    Pan, H.; Zhang, Y.-W. Edge-dependent structural, electronic and magnetic properties of $MoS_2$ Nanoribbons. *J. Mater. Chem.* **2012**, *22*, 7280-7290.

39.    Kresse, G.; Furthmüller, J. Efficient iterative schemes for ab initio total-energy calculations using a plane-wave basis set. *Phys. Rev. B* **1996**, *54*, 11169.

40.    Kresse, G.; Joubert, D.  From ultrasoft pseudopotentials to the projector augmented-wave method. *Phys. Rev. B* **1999**, *59*, 1758.

41.    Liu, J.; Sun, Q.; Kawazoed, Y.; Jena, P. Exfoliating biocompatible ferromagnetic Cr-trihalide monolayers. *Phys. Chem. Chem. Phys.* **2016**, *18*, 8777-8784.

42.    Giannozzi, P.; Andreussi, O.; Brumme, T.; Bunau, O.; Buongiorno Nardelli, M.; Calandra, M.; Car, R.; Cavazzoni, C.; Ceresoli, D.; Cococcioni, M.; Colonna, N.; Carnimeo, I.; Dal Corso, A.; de Gironcoli, S.; Delugas, P.; DiStasio Jr, R. A.; Ferretti, A.; Floris, A.; Fratesi, G.; Fugallo, G.; Gebauer, R.; Gerstmann, U.; Giustino, F.; Gorni, T.; Jia, J.; Kawamura, M.; Ko, H.-Y.; Kokalj, A.; Küçükbenli, E.; Lazzeri, M.; Marsili, M.; Marzari, N.; Mauri, F.; Nguyen, N. L.; Nguyen, H.-V.; Otero-de-la-Roza, A.; Paulatto, L.; Poncé, S.; Rocca, D.; Sabatini, R.; Santra, B.; Schlipf, M.; Seitsonen, A. P.; Smogunov, A.; Timrov, I.; Thonhauser, T.; Umari, P.; Vast, N.; Wu, X.; Baroni, S. Advanced capabilities for materials modelling with Quantum ESPRESSO. *J. Phys.: Condens. Matter* **2017**, *29*, 465901.

Table 1. The magnetic moment (in Bohr magneton $\mu_B$) of atoms in the Z4CrI$_3$ flat nanoribbon calculated with different methods. Atom indexes 1-4 represent Cr atoms, 10-13 for the four edge I atoms and "others" for the other non-edge I atoms, see Figure2(e) for the atoms with the labeled indexes. The "±" in column "others" represents the varying range of the numbers.

| Atom index | 10, 12 | 3 | 1 | 2 | 4 | 11,13 | others |
|---|---|---|---|---|---|---|---|
| LSDA-U | 0.235 | 2.969 | 2.996 | 2.997 | 2.968 | 0.234 | -0.055±0.007 |
| PBE | 0.245 | 3.029 | 3.070 | 3.070 | 3.028 | 0.250 | -0.061±0.006 |
| SCAN | 0.262 | 3.049 | 3.097 | 3.097 | 3.049 | 0.266 | -0.065±0.006 |
| PBE-U | 0.224 | 3.323 | 3.364 | 3.365 | 3.323 | 0.228 | -0.110±0.006 |
| mTASK | 0.225 | 3.589 | 3.632 | 3.634 | 3.589 | 0.229 | -0.213±0.006 |
| TASK | 0.215 | 3.507 | 3.557 | 3.558 | 3.507 | 0.208 | -0.182±0.008 |
| r$^2$SCAN | 0.251 | 3.137 | 3.182 | 3.183 | 3.136 | 0.254 | -0.080±0.007 |

Table 2. The energy difference (in meV/atom) between cases with different directions of the magnetic field for the Z4CrI$_3$ nanoribbon under different tensile and bending, calculated with PBE-U and the spin-orbit coupling (SOC) effects included. The direction of magnetic field is selected by the orientation of the spin quantization axis. The supercell vectors a, b and c are along the directions of axes x, y, and z, respectively. 010 represents the direction of vector b (or axis y), and so on. Orientation 001 is set to be the reference, and the entries for other orientations are calculated with respect to this reference. A positive value represents higher energy than the case of orientation 001. It shows that the three orientations (011, 111 and 101) surrounding the 001 orientation have higher energies and the three orientations (011,111 and 110) surrounding the 010 orientation also have higher energies. However, orientation 100 has higher energy than its three surrounding orientations (110, 111 and 101). This means the two orientations 001 and 010 are locally stable.

| Spin quantization axis | 001 | 010 | 111 | 011 | 100 | 110 | 101 |
|---|---|---|---|---|---|---|---|
| T5 | 0.00 | 0.27 | 0.69 | 0.55 | 0.97 | 0.62 | 0.48 |
| T2 | 0.00 | 0.29 | 0.67 | 0.56 | 0.90 | 0.59 | 0.44 |
| flat | 0.00 | 0.31 | 0.66 | 0.57 | 0.85 | 0.58 | 0.42 |
| R8 | 0.00 | 0.35 | 0.60 | 0.57 | 0.70 | 0.51 | 0.34 |
| R5 | 0.00 | 0.41 | 0.52 | 0.54 | 0.52 | 0.45 | 0.25 |

Table 3. The calculated data for the Z6CrI$_3$ nanoribbon. Data is similarly obtained and organized as in Table 2. It shows that the three orientations (011, 111 and 101) surrounding the 001 orientation have higher energies and the three orientations (011,111 and 110) surrounding the 010 orientation also have higher energies. However, orientation 100 has higher energy than its three surrounding orientations (110, 111 and 101). This means the two orientations 001 and 010 are locally stable, like the case of the Z4CrI$_3$ nanoribbon.

| Spin quantization axis | 001 | 010 | 111 | 011 | 100 | 110 | 101 |
|---|---|---|---|---|---|---|---|
| flat | 0.00 | 0.19 | 0.45 | 0.38 | 0.59 | 0.39 | 0.29 |
| R10 | 0.00 | 0.28 | 0.38 | 0.38 | 0.40 | 0.33 | 0.19 |
| R8 | 0.00 | 0.30 | 0.53 | 0.40 | 0.65 | 0.51 | 0.36 |
| R6 | 0.00 | 0.37 | 0.68 | 0.40 | 0.90 | 0.73 | 0.55 |
| R5 | 0.00 | 0.44 | 0.83 | 0.45 | 1.04 | 0.91 | 0.69 |

Table 4. The calculated data for the Z8CrI$_3$ nanoribbon. Data is similarly obtained and organized as in Table 2. It shows that the three orientations (011, 111 and 101) surrounding the 001 orientation have higher energies. This means that 001 is locally stable. For 010 orientation, from flat to R11, the three orientations (011,111 and 110) surrounding the 010 orientation have higher energies, however, for R7 and R6, orientation 011 has lower energies than the 010 orientation. This means that for small bending curvatures, orientation 010 is still a locally stable direction, however, for larger bending curvatures, orientation 010 is no longer locally stable, and only orientation 001 is the stable one.

| Spin quantization axis | 001 | 010 | 111 | 011 | 100 | 110 | 101 |
|---|---|---|---|---|---|---|---|
| flat | 0.00 | 0.10 | 0.33 | 0.27 | 0.46 | 0.28 | 0.23 |
| R16 | 0.00 | 0.16 | 0.33 | 0.27 | 0.40 | 0.29 | 0.21 |
| R11 | 0.00 | 0.21 | 0.34 | 0.27 | 0.39 | 0.33 | 0.22 |
| R7 | 0.00 | 0.36 | 0.46 | 0.25 | 0.56 | 0.55 | 0.37 |
| R6 | 0.00 | 0.42 | 0.53 | 0.23 | 0.71 | 0.68 | 0.48 |

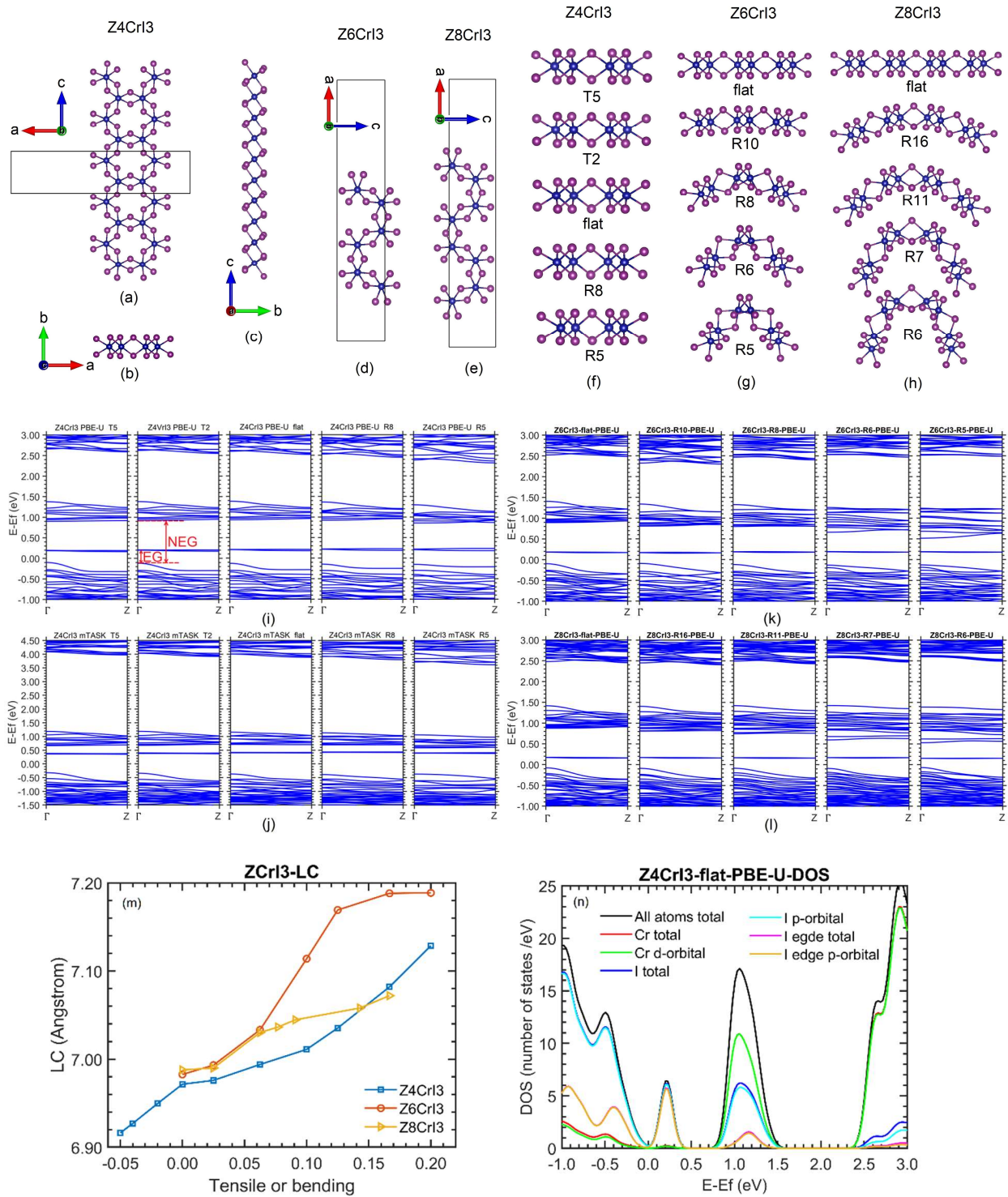

Figure 1. The structures of CrI₃ nanoribbons, calculated band structures, supercell lattice constant, and density of states. The supercell's vectors a, b and c are orientated along axes x, y, and z, respectively. The structure of Z4CrI₃ is in (a) (top view), in (b) (side view) and in (c) (side view). Blue balls represent Cr atoms and purple ones for I atoms. The structure of Z6CrI₃ is in (d) and Z8CrI₃ in (e). The boxes in (a), (d) and (e) indicate the periodic unit of the nanoribbon. (f)-(h) show the relaxed structures of Z4CrI₃,

Z6CrI$_3$, and Z8CrI$_3$ nanoribbons under different bending or tensile conditions, where T5 (T2) represents the tensile strain of 5% (2%) along the width direction (vector a direction) with respect to the monolayer structure, and R8 represents the bending curvature radius R = 8 Å and so on for others. Note that after relaxation, all Z4CrI$_3$ nanoribbons are almost flat. (i) shows the band structures of Z4CrI$_3$ nanoribbons under different bending or tensile strains (from left to right: tensile 5%, tensile 2%, flat with no bending or tensile strain, bending R = 8 Å, and R = 5 Å) with the PBE-U calculation. EG (edge gap) represents the energy gap between the edge band and the valence top at Γ, while NEG (non-edge gap) represents that between the lowest non-edge conduction band and the valence top at Γ, as depicted in one of the panels in (i). (j) shows the band structure of Z4CrI$_3$ from the mTASK calculation. (k) and (l) show the PBE-U band structures of Z6CrI$_3$ and Z8CrI$_3$ nanoribbons, respectively. The changing trend of the length of the supercell's vector c with bending or tensile strains for the three nanoribbons is in (m). The partial density of states (DOS) of flat Z4CrI$_3$ is in (n).

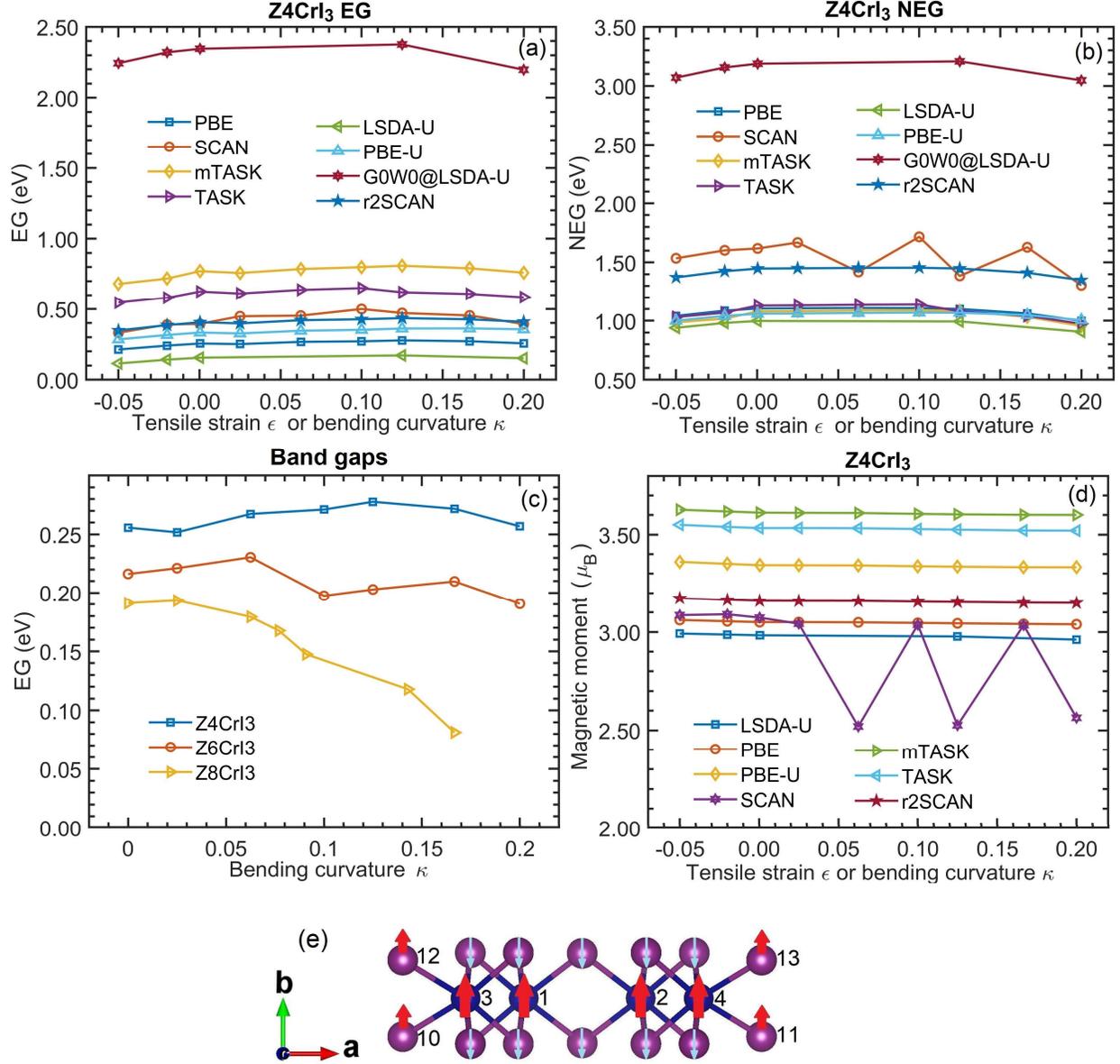

Figure 2. The edge band gap (EG) and non-edge band gap (NEG) of CrI$_3$ nanoribbons and the average magnetic moment of Cr atoms. EG and NEG of the Z4CrI$_3$ nanoribbon under different tensile or bending calculated with PBE, SCAN, r$^2$SCAN, PBE-U (with U=2.65 eV for Cr atoms[41]), mTASK, TASK, LSDA-U (with U = 1.5 eV and J = 0.5 eV for Cr atoms[10]), and G$_0$W$_0$@LSDA-U are in (a) and (b), respectively. The PBE-U results of EGs of Z4CrI$_3$, Z6CrI$_3$, and Z8CrI$_3$ with bending curvatures are in (c). The average magnetic moment of the four Cr atoms in the cell of Z4CrI$_3$ nanoribbon with different tensile strains and bending curvatures is in (d). The schematic of the orientation of magnetic moment of atoms is in (e). Blue balls represent Cr atoms and purple ones for I atoms. The bending curvature is $\kappa = 1/R$, with a unit of 1/Å, where R is the bending curvature radius. The tensile strain $\epsilon$ along the ribbon width direction is represented as a negative number, with -0.05 (-0.02) representing 5% (2%) tensile strain.

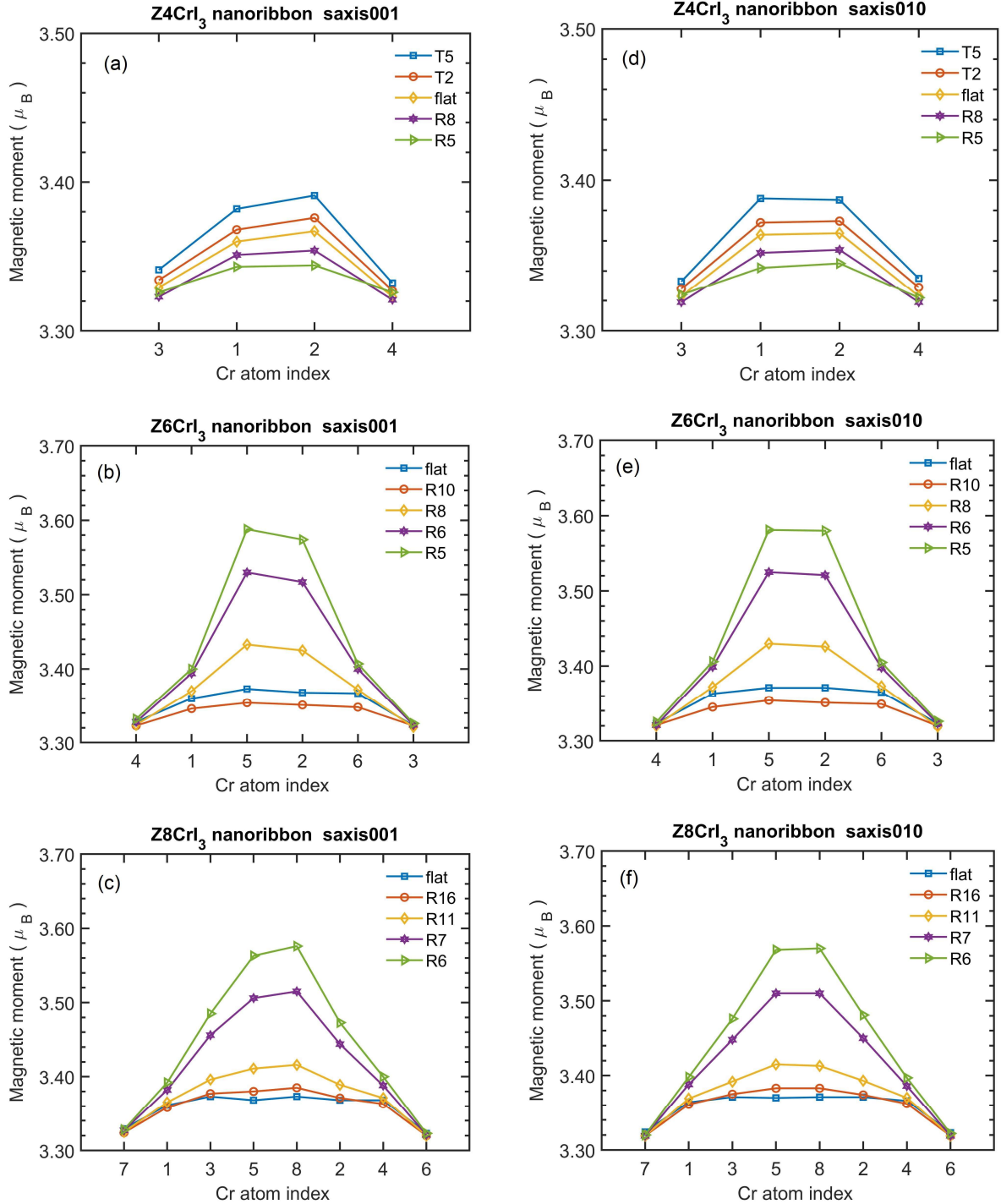

Figure 3. The magnetic moments on Cr atoms in the Z4CrI$_3$, Z6CrI$_3$ and Z8CrI$_3$ nanoribbons as a function of the bending or tensile strains. (a)-(c) are for the spin quantization axis along the periodical z axis (in-plane) for Z4CrI$_3$, Z6CrI$_3$ and Z8CrI$_3$ nanoribbons, respectively. (d)-(f) are for the spin quantization axis along the y axis (out-of-plane). The atom indexes shown in the middle region of the plots are for Cr atoms located in the middle region of the nanoribbon, while the most left and most right ones are located near the two edges of the ribbon.

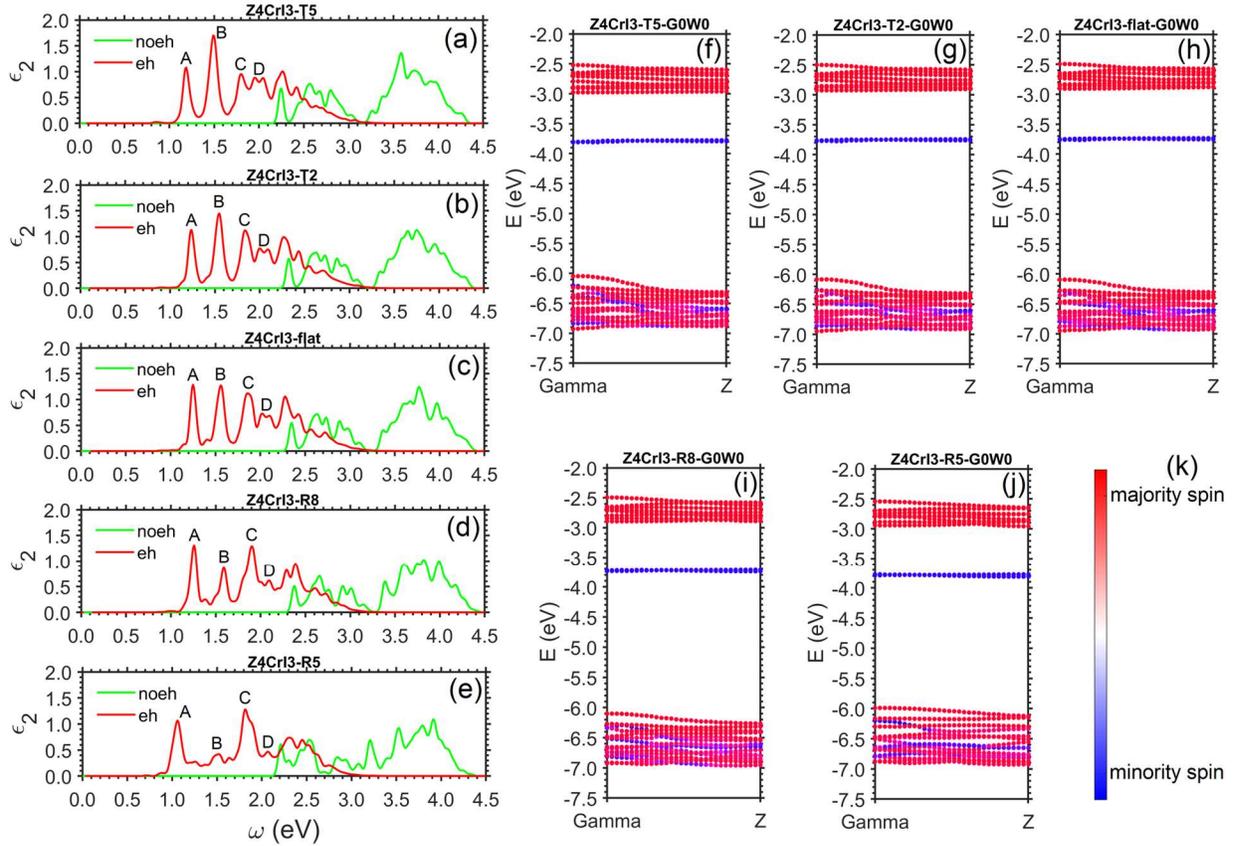

Figure 4. The optical absorption spectra and $G_0W_0$ quasiparticle band structures of nanoribbon Z4CrI$_3$ under different tensile strains or bending curvatures. The optical spectra of tensile strain 5% (T5) is in (a), 2% (T2) is in (b), flat ribbon is in (c), bending curvature $\kappa = 0.125$ /Å (or R = 8 Å) is in (d), and $\kappa = 0.20$ /Å (or R = 5 Å) is in (e). All curves are calculated with a Gaussian broadening of 28 meV. "eh" and "noeh" in legends stand for with electron-hole interactions (GW-BSE method) and without electron-hole interactions (GW-RPA method), respectively. From (f)-(j), the spin resolved $G_0W_0$@LSDA-U quasiparticle band structures are for T5, T2, flat, R8 and R5, respectively. (k) is the color bar for the spin polarization of the states. The spin polarization is calculated as $\langle \psi_{n,k} | \sigma_z | \psi_{n,k} \rangle$, where $\psi_{n,k}$ is the wavefunction and $\sigma_z$ the Pauli matrix.

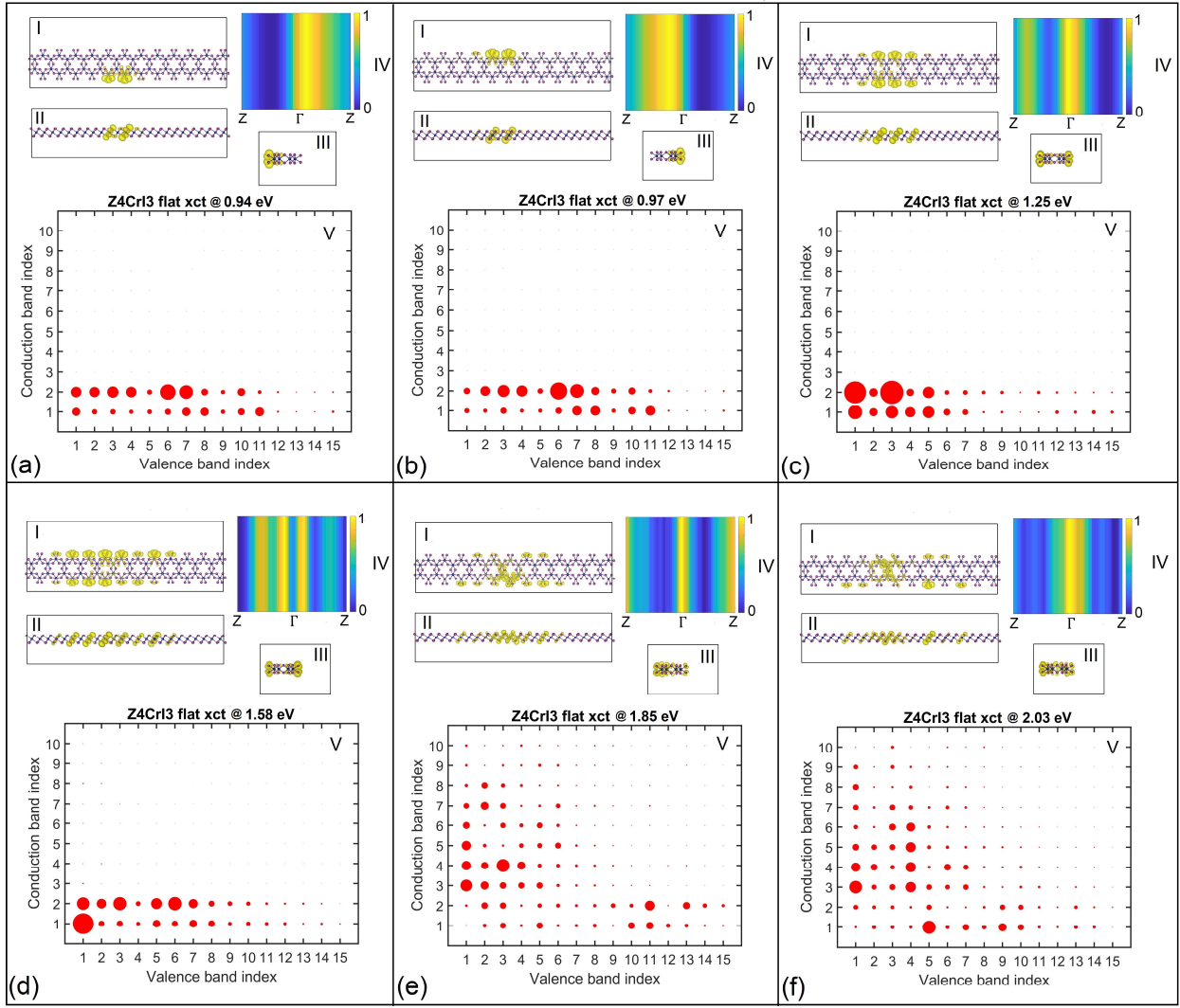

Figure 5. The wavefunction and the electron-band and hole-band composition of excitons at energy of (a) 0.94 eV, (b) 0.97 eV, (c) 1.25 eV, (d) 1.58 eV, (e) 1.85 eV, and (f) 2.03 eV of the Z4CrI$_3$ flat nanoribbon. For each exciton, the subplots I, II and III are the squared modulus of exciton wavefunction in real space in its top and side views. The hole is located about the sixth Cr atom in the third Cr atom row in subplot I. Blue balls represent Cr atoms and purple ones for I atoms. Subplot IV is the band-summed envelope function (arbitrary unit) of exciton, $\sum_{v,c} |A_{v,c}(k)|^2$, representing the distribution of excitons in k-space. The contributing hole (valence) and electron (conduction) bands for each exciton is in subplot V. The valence (conduction) band index is counted downwards (upwards) from the Fermi level. The spot size in subplot V is proportional to $\sum_k |A_{v,c}(k)|^2$ and represents the contributing weight from the v-c pair.

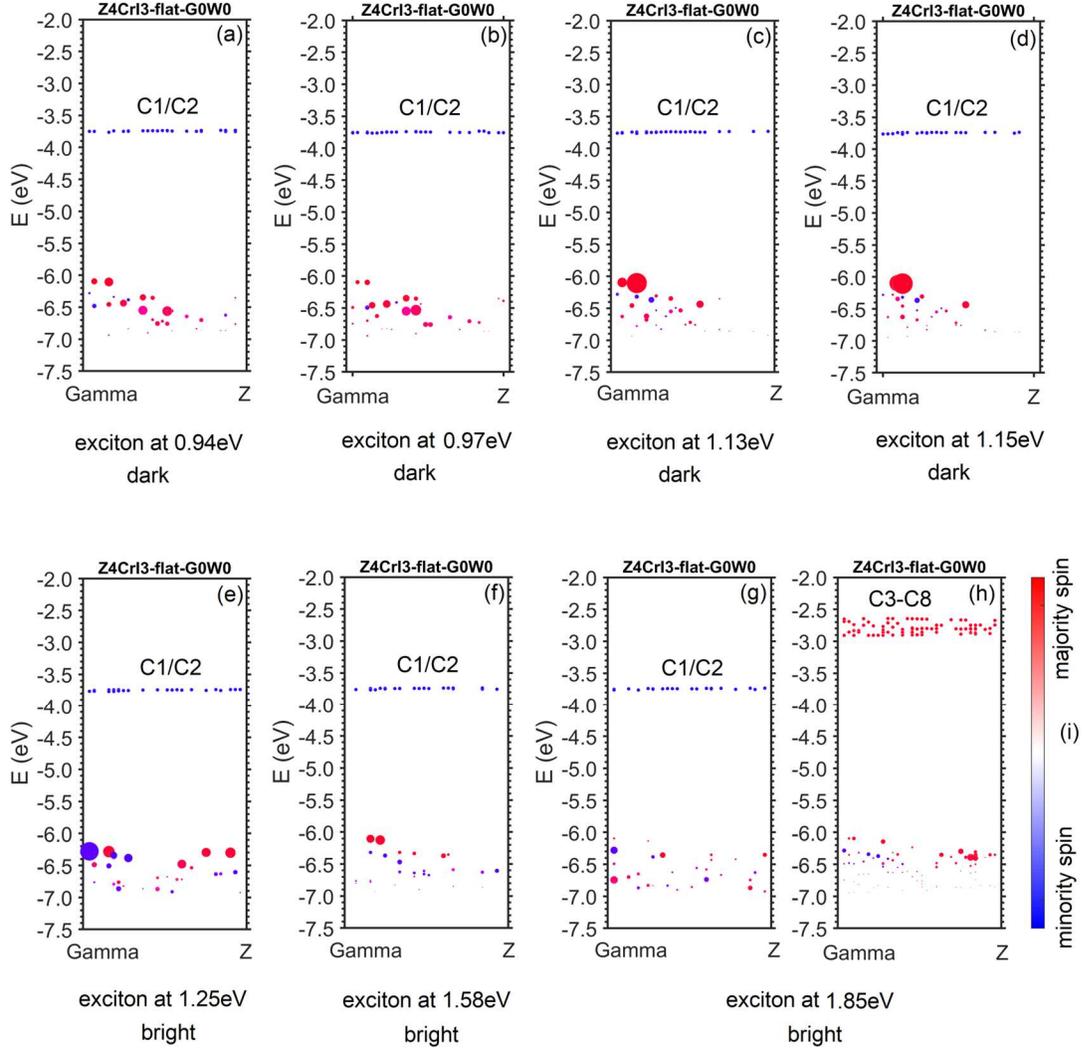

Figure 6. The spin polarization resolved valence (or hole) and conduction (or electron) states involved in excitons in the Z4CrI$_3$ flat nanoribbon. The spot size in the valence states in each plot is proportional to the modulus square of the exciton envelope function $\left|A_{v,c}(k)\right|^2$. (a)-(d) are the dark excitons at 0.94 eV, 0.97 eV, 1.13 eV and 1.15 eV, respectively. As can be seen, most of the states in the valence bands are of majority spin (red) and all the states in the edge conduction bands are of minority spin (blue), indicating the formation of the dark exciton mainly from unlike-spin polarization. (e) and (f) show the bright excitons at 1.25 and 1.58 eV, respectively. Both (g) and (h) are for the bright exciton at 1.85 eV with (g) for transitions to edge bands C1/C2 and (h) for transitions to conduction bands C3-C8. As can be seen, there are significant contributions from the minority spin (blue) polarization in the valence states in (e)-(g), and a significant contribution from the majority spin (red) polarization in the valence states in (h), indicating a mixed spin configuration in bright excitons with both unlike-spin and like-spin polarization configurations. (i) is the color bar.

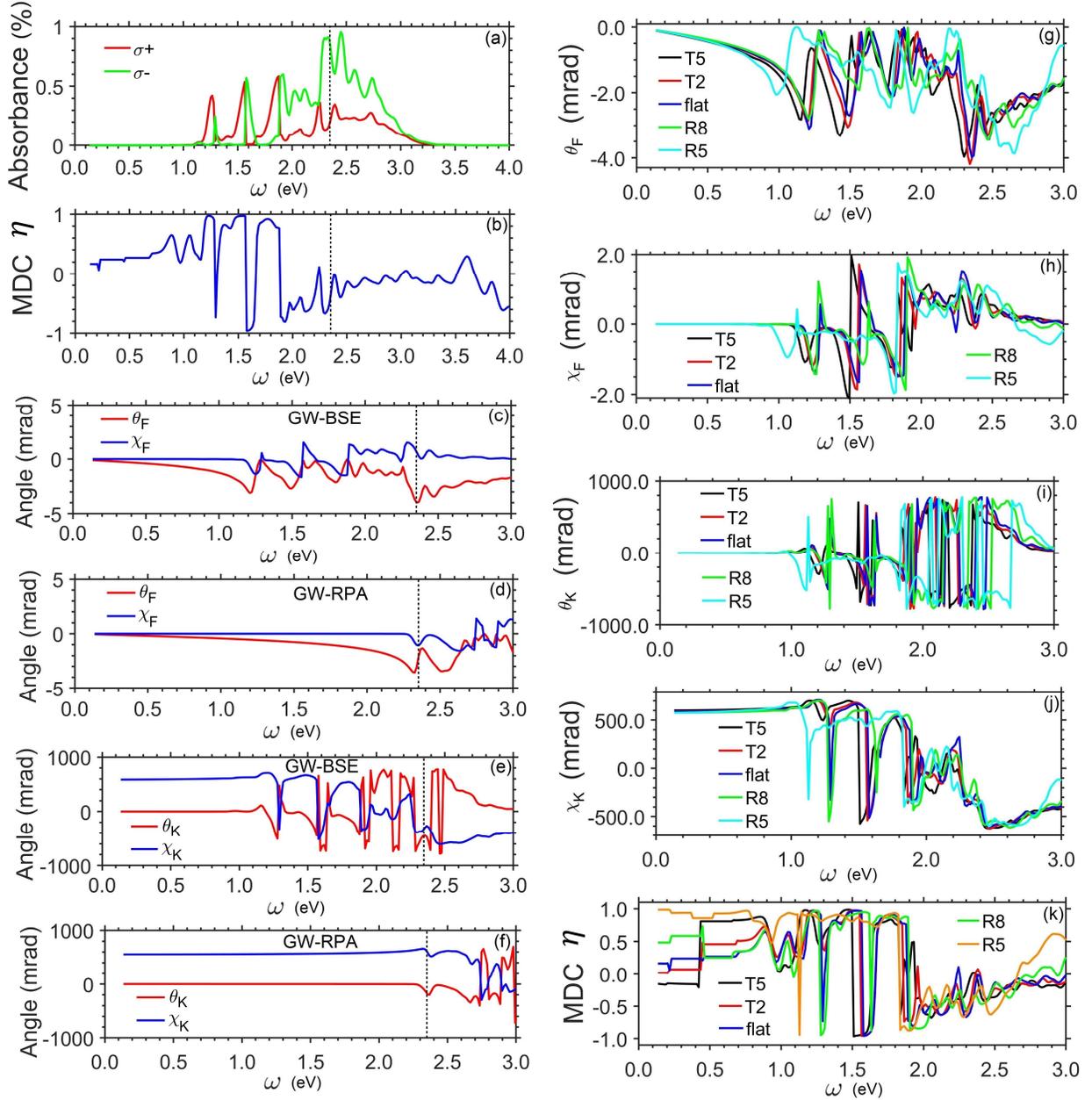

Figure 7. Magneto-optical properties of the Z4CrI$_3$ nanoribbon. The absorbance spectra of the left ($\sigma^+$) and right ($\sigma^-$) elliptically polarized modes (EPMs) is in (a) and its magnetic dichroism (MDC) contrast $\eta$ is in (b). The transmitted Faraday rotation angle $\theta_F$ and ellipticity $\chi_F$ with e-h interaction (GW-BSE) versus photon energy is in (c). The same quantities as in (c) without e-h interaction (GW-RPA) is in (d). The reflective magneto-optical (MO) Kerr rotation $\theta_K$ and ellipticity $\chi_K$ with e-h interactions (GW-BSE) versus photon energy is in (e) and without e-h interactions (GW-RPA) in (f). The dashed vertical lines in (a)-(f) denote the quasiparticle band gap. The Faraday $\theta_F$ under different tensile strains and bending curvatures is in (g), and the Faraday $\chi_F$ under different tensile strains and bending curvatures is in (h). The MO Kerr $\theta_K$ and $\chi_K$ under different tensile strains and bending curvatures is in (i) and (j),

respectively. The MDC contrast under different tensile strains and bending curvatures is in (k). The GW-BSE results are shown in (g)-(k).